\documentclass[letterpaper,10pt]{article} 

\usepackage{opticameet3} 
\usepackage{multicol} 
\usepackage{silence}
\usepackage{romannum}
\usepackage[font=footnotesize,labelfont=bf]{caption}

\newcommand\authormark[1]{\textsuperscript{#1}}

\usepackage{amsmath,amssymb}
\usepackage[colorlinks=true,bookmarks=false,citecolor=blue,urlcolor=blue]{hyperref} 

\begin{document}



\title{Net 835-Gb/s/$\lambda$ Carrier- and LO-Free 100-km Transmission Using Channel-Aware Phase Retrieval Reception}

\vspace{-16pt} 
\copyrightyear{2024}
\author{
    Hanzi~Huang\authormark{1,2},
    Haoshuo~Chen\authormark{1,*},
    Qian~Hu\authormark{1},
    Di~Che\authormark{1},\\
    Yetian~Huang\authormark{1,2},
    Brian~Stern\authormark{1},
    Nicolas~K.~Fontaine\authormark{1},
    Mikael~Mazur\authormark{1},\\
    Lauren~Dallachiesa\authormark{1},
    Roland~Ryf\authormark{1},
    Zhengxuan~Li\authormark{2},
    and Yingxiong~Song\authormark{2}
    }

\address{\authormark{1} Nokia Bell Labs, 600 Mountain Ave, Murray Hill, NJ 07974, USA\\
\authormark{2} Key Laboratory of Specialty Fiber Optics and Optical Access Networks, Shanghai University, 200444 Shanghai, China\\}
\vspace{-3pt} 
\email{\authormark{*}haoshuo.chen@nokia-bell-labs.com} 

\vspace{-18pt} 

\begin{abstract}
We experimentally demonstrate the first carrier- and LO-free 800G/$\lambda$ receiver \mbox{enabling} direct \mbox{compatibility} with standard coherent transmitters via phase retrieval, achieving net \mbox{835-Gb/s} \mbox{transmission} over 100-km SMF and record 8.27-b/s/Hz net optical spectral \mbox{efficiency}.
\end{abstract}

\vspace{-4pt}

\section{Introduction}
\begin{figure}[b!]
	\vspace{-18pt}
	\centering
	\includegraphics[width=\hsize]{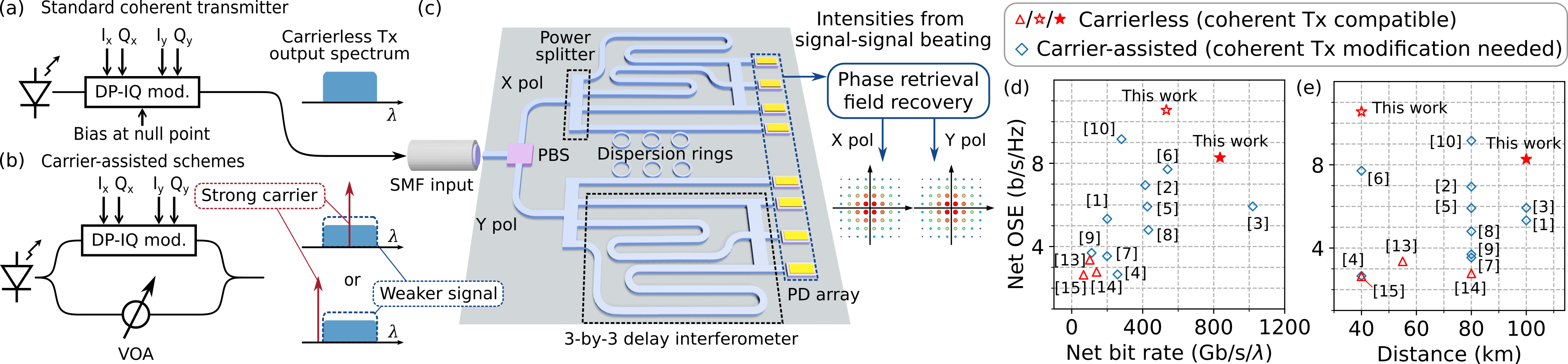}
	\setlength{\abovecaptionskip}{-0.4cm}
	\setlength{\belowcaptionskip}{-0.7cm}
	\vspace{4pt}
	\caption{
		Schematics of (a) a standard coherent transmitter, and (b) a carrier-assisted DP-IQ modulation transmitter. (c) Image of an integrated dual-polarization PR receiver. Net OSE versus (d) net bit rate and (e) link distance of reported transmission experiments using intensity detection and complex-field reconstruction methods.
	}
	\label{fig1}
\end{figure}

\vspace{-4pt}
With the 400ZR standard being implemented in recent years, adopting coherent optics to scale the capacity to 800~Gb/s per lane and beyond is being widely considered.
Notably, as the single-channel rate increases, coherent technology is progressively sinking to shorter distances and broadening its application scenarios.
During this transition, although the cost of coherent transceivers has significantly decreased due to advancements in photonics integration, the stringent requirements for stable local oscillator (LO) lasers and sophisticated remote wavelength alignment/management mechanisms pose challenges in implementing coherent solutions to more widespread, cost-sensitive transmission links.
Addressing these challenges necessitates the development of a direct detection-based receiver (Rx) that is compatible with the standard coherent transmitter structure, as illustrated in Fig.~\ref{fig1}(a). 
This approach, essentially leading to a LO-free coherent Rx, is anticipated to facilitate cost-effective and highly flexible capacity scaling for short-reach applications.
To achieve this goal, an intensively studied approach is integrating an auxiliary optical carrier at the transmitter (Tx), as depicted in Fig.~\ref{fig1}(b), and enable channel linearization by detecting signal-carrier beating components at Rx, namely as carrier-assisted schemes.
Techniques such as Kramers-Kronig Rx~\cite{KK, pol_track_Jones_space_KK}, Stokes-vector Rx~\cite{Stokes},  asymmetric self-coherent detection (ASCD)~\cite{ASCD_silicon_photonic}, carrier-assisted
differential detection (CADD)~\cite{CADD_silicon_photonic_ECOC}, and many other variants~\cite{self_homodyne_silicon_photonic, deep_learning_xingfeng, son_thai_le_SSB, CAPR, DP_CAPR} fall into this category.
Carrier-assisted schemes rely on a sufficient carrier-to-signal power ratio (CSPR) to ensure signal detection.
This fundamental difference sets them apart from the anticipated LO-free coherent Rx since only a small portion of the amplifier's output power is allocated for amplifying the information-bearing signal, thereby compromising the transmission efficiency and making them prone to operating under optical signal-to-noise ratio (OSNR)-limited conditions.
While boosting the launch power per channel may alleviate this issue, it necessitates integrating more advanced optical amplifiers with much higher output power into optical transceiver modules and leads to extra performance degradation due to fiber nonlinearity, especially in WDM scenario, considering each channel occupies a strong carrier~\cite{son_thai_le_SSB}.

To eliminate both carrier and LO, a carrierless approach known as the phase retrieval (PR) Rx has been developed~\cite{PR_Haoshuo,2D_PD_PR,SpaceTime_PR,JLT_distortion_PR,PR_weight_decision}.
This strategy does not require any modification to the standard coherent transmitter, and its Rx optical front-end is fully compatible with photonic integration~\cite{Brian_PR}.
Fig.~\ref{fig1}(c) illustrates the rendered 3D image of an integrated dual-polarization PR Rx.
Each polarization of the input signal is received by a space-time diversity PR Rx~\cite{SpaceTime_PR}. 
In the dispersive and interferometry branches, the phase information of the modulated signal is converted into intensity fluctuations via a deliberate strong symbol mixing effect. 
During the subsequent digital signal processing (DSP) procedure, the signal phase is deduced from these measured intensity waveforms with the aid of an iterative field reconstruction algorithm.
As a result, the PR Rx can be designed as fully compatible with a standard coherent transmitter, functioning in a colorless intensity detection manner and obviating the need for any LO or reference carrier.
With the aid of a newly designed channel impairment estimation and compensation scheme for PR reception, this paper experimentally demonstrates 100-GBaud dual-polarization (DP) probabilistically shaped (PS) 64-quadrature amplitude modulation (QAM) transmission over 100-km single-mode fiber (SMF) without resorting to any LO or carrier, achieving a net rate of 835~Gb/s and an 8.27-b/s/Hz net optical spectral efficiency (OSE) record, as marked by the solid red star in Figs.~\ref{fig1}(d) and (e).
50-GBaud DP-PS256QAM transmission with 7.2-bit/symbol entropy over 40-km SMF with net 533-Gb/s bit rate and 10.55-b/s/Hz net OSE is also demonstrated, as marked by the hollow red star.
The results show that our approach achieves the highest net bit rate among carrierless schemes, as well as the highest net OSE compared with other reported carrier-assisted schemes of similar net bit rates.
In Fig.~\ref{fig1}(e), the net OSE record is achieved under a single span transmission of 100~km, a distance at which decreased OSNR starts to limit the net OSE for all the schemes.
This highlights the advantage of the carrierless PR scheme in operating under OSNR-constrained links by saving extra optical power from transmitting the carrier.
\begin{figure}[b!]
	\vspace{-14pt}
	\centering
	\includegraphics[width=\hsize]{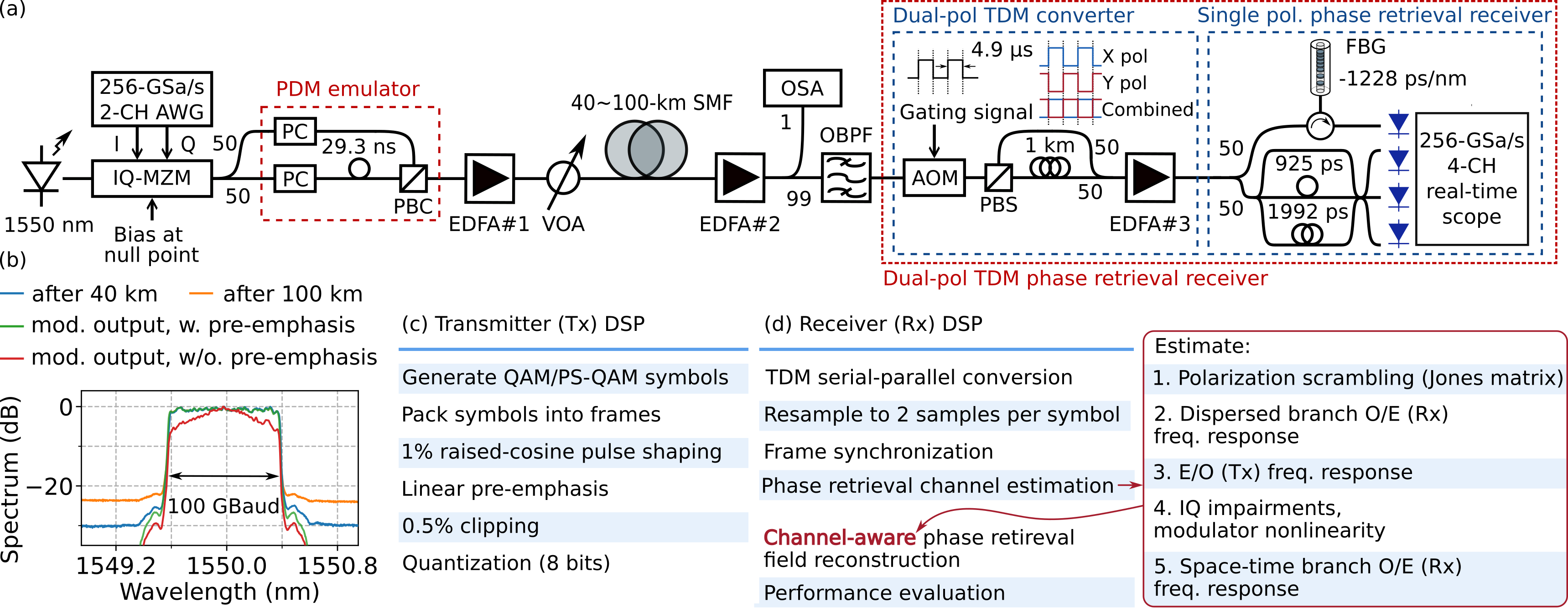}
	\setlength{\abovecaptionskip}{-0.4cm}
	\setlength{\belowcaptionskip}{-0.6cm}
	\vspace{4pt}
	\caption{
		(a) Experimental setup. (b) Measured optical spectra. (c) Tx and (d) Rx DSP procedures.
	}
	\label{fig2}
\end{figure}
\vspace{-8pt}
\section{Experimental Setup}
\vspace{-4pt}

\begin{figure}[t!]
	\centering
	\includegraphics[width=\hsize]{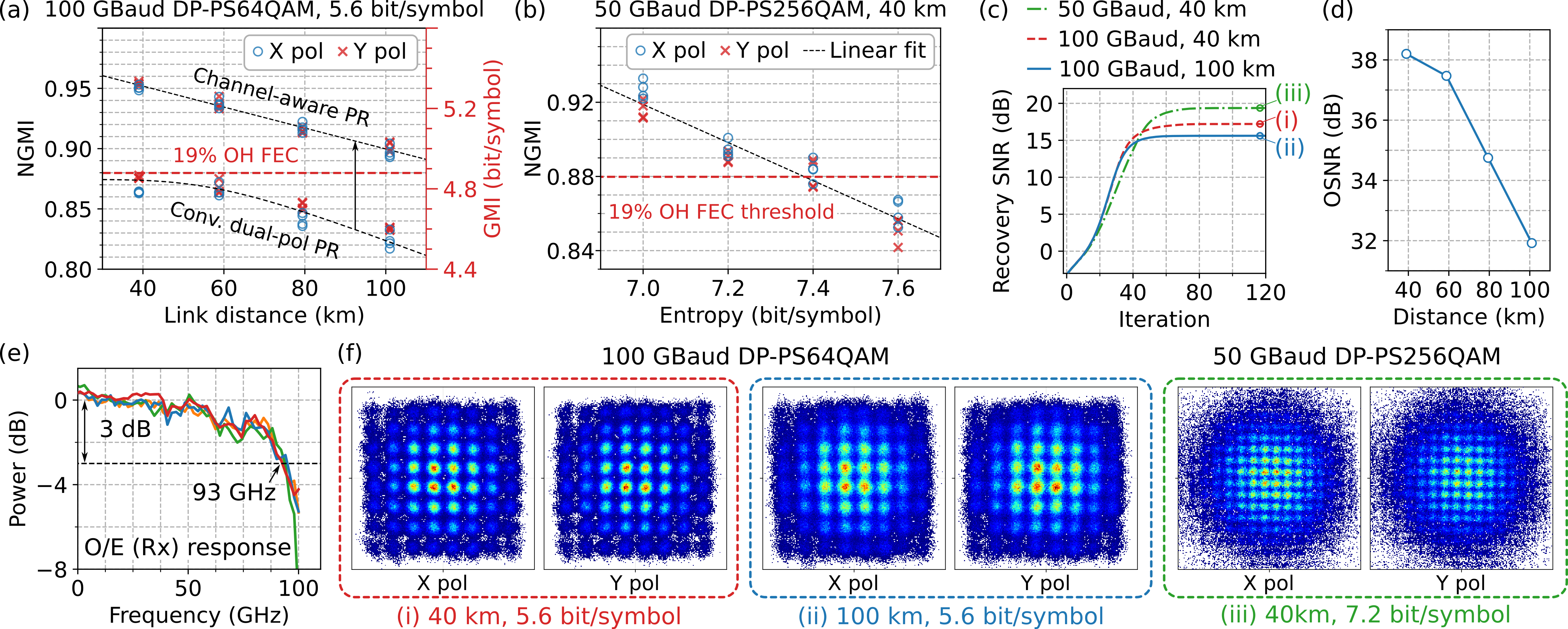}
	\setlength{\abovecaptionskip}{-0.4cm}
	\setlength{\belowcaptionskip}{-0.6cm}
	\vspace{4pt}
	\caption{
        (a) NGMI and GMI versus link distance of 100-GBaud 5.6-bit/symbol DP-PS64QAM. (b) NGMI versus entropy of 50-GBaud DP-PS256QAM over 40-km SSMF. (c) Recovery SNRs versus iteration number. (d) Measured OSNR versus link distance. (e) Estimated receiver optical-electrical (O/E) frequency power response. (f) Recovered constellations.
	}
	\label{fig3}
\end{figure}

Figure~\ref{fig2}(a) presents the experimental setup of the PR-based system receiving DP PS-QAM signal.
At the transmitter side, an external-cavity laser (ECL) with a linewidth of 100-kHz at 1550~nm is modulated by an IQ-Mach-Zehnder modulator (MZM) to produce the modulated signal with a roll-off factor of 1\%.
The electrical signal is generated by a 2-channel 256-GSa/s arbitrary waveform generator (AWG) (Keysight M8199B). 
After modulation, the polarization division multiplexed (PDM) signal is emulated by polarization-combining the single-polarization signal with its decorrelated copy after 29.3-ns (5.86-m) delay.
The PDM signal is then pre-amplified to 4~dBm by an erbium-doped fiber amplifier (EDFA) before launching into the transmission SMF.
At the Rx, another EDFA is used to re-boost the received optical power, with its output connected to a 99:1 splitter, with the 1\% branch connected to an optical spectrum analyzer (OSA) for OSNR measurement.
An optical band-pass filter (OBPF) is used to filter out-of-band noise with a bandwidth of $\sim$1~nm.
To realize the reception of the 8-channel signal using one 4-channel real-time oscilloscope (RTO), a time division multiplexing (TDM) converter is introduced to map the two received polarization signals into two sequential time frames.
An acousto-optic modulator (AOM) is driven with a 10-\textmu s period and 49\% duty cycle.
Following this, the PDM signal undergoes polarization demultiplexing and is recombined into a serial TDM signal through a 1-km delay SMF and a 50:50 coupler.
An additional EDFA (marked as \#3) compensates for the losses incurred by the TDM converter, including the \mbox{2.8-dB} intrinsic loss from the AOM and 3.1-dB loss from the gating modulation, which can be omitted if the TDM scheme is not utilized.
Subsequently, the TDM signal is sent into a single polarization space-time diversity PR Rx.
The dispersive element used in the PR Rx is a fiber Bragg grating (FBG)-based dispersion compensation module with an estimated chromatic dispersion of -1228~ps/nm at the operating wavelength.
The measured time delays of the two interfering tributaries are around 93 and 199 symbol periods compared to the shortest branch.
The received optical signals are detected by four 100-GHz photodiodes with average received power around 5~dBm to fully suppress the thermal noise, and the detected electrical signals are captured by a 256~GSa/s RTO (Keysight UXR1004A).
Transmission performance is measured under different single mode fiber (SMF) span lengths from 40 to 100 km.
Fig.~\ref{fig2}(b) shows the measured optical spectra at both the Tx and Rx.

Fig.~\ref{fig2}(c) and (d) illustrate the offline transceiver DSP flows.
We enhance the performance of PR field reconstruction with channel awareness by including Tx electrical-optical (E/O) complex-valued and Rx optical-electrical (O/E) real-valued frequency response, IQ imbalance, modulator nonlinearity and polarization scrambling effect into a modified Gerchberg-Saxton (GS) algorithm. 
The Rx frequency response distortion is compensated using feed-forward equalizers (FFEs) with 101 half-symbol-spaced taps.
In the reconstruction stage, the forward GS propagation is designed to mirror the real-world physical model of light propagation considering field-based Tx distortion, and polarization rotation as well as chromatic dispersion effect from the fiber channel. 
The backward GS propagation is derived as the reverse of the forward propagation process to compensate the distortion.
To enhance convergence, frequency bandwidth constraint and selective phase reset operations are incorporated~\cite{PR_Haoshuo,PR_reset}.
Evenly distributed 10\% pilot symbols is applied to eliminate the global phase ambiguity.
Various channel impairments and polarization rotation effects are estimated through a PR-based channel estimation using a training sequence with 8192 uniform 64QAM symbols by minimizing errors between the distorted forward propagated training sequence's intensity and the measured intensity waveform~\cite{JLT_distortion_PR}.
Most channel impairments are static and require estimation only once.
The polarization scrambling effect, represented by the Jones matrix, can be updated at the millisecond level in practical implementation.
\vspace{-10pt}
\section{Experimental Results}
\vspace{-4pt}

After recovering the complex-valued symbols, the system performance is evaluated by generalized mutual information (GMI) and normalized GMI (NGMI), which predict the post-FEC performance reliably.
Fig.~\ref{fig3}(a) shows the NGMI and GMI results versus transmission distance of 100~GBaud DP-PS64QAM signal with 5.6-bit/symbol entropy at each polarization after 100 reconstruction iterations.
Notable performance enhancements can be observed upon incorporating the estimated channel state information into the field reconstruction process.
A NGMI threshold of 0.8798 is given, corresponding to 19.02\% FEC overhead~\cite{FEC,vivan}.
Four independent measurements are taken to demonstrate its performance stability over received polarization state variations and over 62 thousand recovered symbols are used to calculate each point in Fig.~\ref{fig3}(a).
To investigate the performance limit under minimal Rx frequency and OSNR impairments, Fig.~\ref{fig3}(b) shows the NGMI versus different entropy of 50~GBaud DP-PS256QAM signals over \mbox{40-km} standard SMF transmission.
Fig.~\ref{fig3}(c) shows the averaged recovery SNR results versus iteration number under three transmission cases.
It can be seen that 80 iterations are sufficient for the algorithm to converge.
The final SNRs after full convergence for cases (\romannum{1}$\sim$\romannum{3}) are 17.21, 15.61 and 19.37~dB, respectively.
Fig.~\ref{fig3}(d) depicted the measured OSNRs versus transmission distances.
Fig.~\ref{fig3}(e) shows the estimated normalized Rx power frequency response of the four channels, with the worst 3-dB roll-off point at 93~GHz.
Fig.~\ref{fig3}(f) show the recovered constellations of the recovered payload symbols.
With 100-GBaud DP-PS64QAM, the net bit rate is calculated as \mbox{2$\times$100$\times$0.9$\times$[5.6-19.02\%/(1+19.02\%)$\times$6]=835.4~Gbit/s~\cite{PCS}}, achieving 8.27-b/s/Hz net OSE.
\mbox{533-Gb/s} net bit rate and \mbox{10.55-b/s/Hz} net OSE are achieved with 50-GBaud \mbox{7.2-bit/symbol} DP-PS256QAM.

\vspace{-10pt}
\section{Conclusions}
\vspace{-4pt}
We demonstrate net 835-Gb/s/$\lambda$ transmission over 100 km with 100-GBaud DP PS-64QAM at 5.6-bit/symbol entropy via a 100-GHz PR Rx, which is the first reported 800G/$\lambda$ intensity detection-based Rx directly supporting standard coherent transmitters.
It shows that PR can be a promising solution for realizing a cost-effective LO-free coherent Rx for future data center and metropolitan network interconnects.

\scriptsize
\vspace{1pt}
\noindent
This work was supported in part by the National Key Research and Development Program of China (2021YFB2900801); Science and Technology Commission of Shanghai Municipality (22511100902, 22511100502); 111 project (D20031).

\end{document}